\documentclass{Interspeech}

\usepackage{tikz}
\usepackage{subcaption}

\usepackage{multirow}
\interspeechcameraready 

\title{Objective and Subjective Evaluation of Diffusion-Based Speech \\ Enhancement for Dysarthric Speech}

\author[affiliation={1}]{Dimme}{de Groot}
\author[affiliation={1,2}]{Tanvina}{Patel}
\author[affiliation={1,3}]{Devendra}{Kayande}
\author[affiliation={1}]{Odette}{Scharenborg}
\author[affiliation={1}]{Zhengjun}{Yue}

\affiliation{Multimedia Computing Group}{Delft University of Technology}{The Netherlands}
\affiliation{Dept. of Plastic and Reconstructive Surgery}{Erasmus Medical Centre (EMC)}{The Netherlands}
\affiliation{Dept. of Electronics and Communication}{Indian Inst. of Information Technology Allahabad}{India}
\email{d.c.c.j.degroot@tudelft.nl}


\usepackage{comment}

\begin{document}

\maketitle

\begin{abstract}
Dysarthric speech poses significant challenges for automatic speech recognition (ASR) systems due to its high variability and reduced intelligibility.
In this work we explore the use of diffusion models for dysarthric speech enhancement, which is based on the hypothesis that using diffusion-based speech enhancement moves the distribution of dysarthric speech closer to that of typical speech, which could potentially improve dysarthric speech recognition performance. We assess the effect of two diffusion-based and one signal-processing-based speech enhancement algorithms on intelligibility and speech quality of two English dysarthric speech corpora. We applied speech enhancement to both typical and dysarthric speech  and  evaluate the ASR performance using Whisper-Turbo, and the subjective and objective speech quality of the original and enhanced dysarthric speech. We also fine-tuned Whisper-Turbo on the enhanced speech to assess its impact on recognition performance.
\end{abstract}

\section{Introduction}
\label{sec:Intro}
Speech is a fundamental mode of communication in our daily interactions. However, for individuals with dysarthria, communication through speech is challenging. Dysarthria is a motor speech disorder resulting from neuro-motor conditions such as cerebral palsy, amyotrophic lateral sclerosis (ALS), or Parkinson's disease, which hinders speech production \cite{kent2000research}. Due to the weakness or paralysis in the vocal organs \cite{joseph2012motor, Enderby2013}, dysarthric speech can have reduced articulation precision, slower speech rate, and variations in pitch and loudness \cite{theodoros1994perceptual, darley1969clusters}. 
The distinctive characteristics of dysarthric speech pose barriers for effective human-human and human-machine interaction. Although speech technology could potentially alleviate these problems, existing speech technology solutions often fall short when applied to dysarthric speech. The limited accessibility and usability of these systems for individuals with dysarthria highlights the need for targeted advancements in speech technology.

Recent advancements in self-supervised models and automatic speech recognition (ASR) models trained on diverse multilingual and multitask speech data have made it possible to improve the ASR performance of dysarthric speech - through fine-tuning \cite{Leivaditi2024, Zheng2024}, domain adaptation techniques \cite{Woszczyk2020}, 
and data augmentation \cite{Harvill2021, Leung2024, Wang2024}.
Moreover, front-end processing techniques have successfully been applied to dysarthric speech for downstream speech tasks \cite{vinotha2024enhancing,Reszka2024}.
One of the pre-processing techniques is speech enhancement (SE), which has been used in therapy tools to improve the quality of dysarthric speech \cite{Sivaram2017}.
Dysarthric SE using  convolutional neural networks (CNN) \cite{wang2022dysarthric}, contrastive learning reconstruction \cite{fatemeh2024enhancement}, speech enhancement generative adversarial networks (SEGANs) \cite{vinotha2024enhancing}, has shown quantitative (in terms of  word error rates (WER) for ASR) and qualitative (mean opinion score (MOS)) improvements of the dysarthric speech. 

A new class of SE algorithms is that of diffusion-based generative models \cite{Lemercier2024}. In \cite{Reszka2024}, it was hypothesised that using generative diffusion-based SE on dysarthric speech would move the distribution of the dysarthric speech closer to that of typical speech by removing acoustic dysarthric speech markers. This is indeed what they found in a dysarthric speech detection task. 
In this work, we investigate the hypothesis that this removal of acoustic dysarthric speech markers by generative diffusion-based SE leads to improved dysarthric speech recognition by state-of-the-art ASR, which are typically trained on typical speech. Specifically, we investigate whether diffusion-based SE affects the quality and intelligibility of dysarthric speech across different severity levels.
We consider two generative diffusion-based SE methods and compare it to a baseline signal-processing based SE method using two commonly used English dysarthric speech datasets: UASpeech \cite{Kim2008} and TORGO \cite{rudzicz2012torgo}. As generative SE methods, we use SGMSE \cite{Richter2023}, which was used for dysarthric speech detection in \cite{Reszka2024}, and StoRM, which has been shown to outperform a lightweight version of SGMSE in terms of ASR performance \cite{Lemercier2023}. As the baseline method, we use Noisereduce \cite{Sainburg2020}, which has previously been used to remove stationary noise in UASpeech \cite{Harvill2021, Illa2021}. 

The intelligibility of the enhanced dysarthric speech is evaluated by an ASR experiment with Whisper-Turbo \cite{whisper_turbo}. In a second ASR experiment, we investigate whether enhancing dysarthric speech affects the effectiveness of fine-tuning (FT). The quality of the enhanced dysarthric speech is evaluated objectively and subjectively. Objectively, the quality of the enhanced speech is measured with Deep-noise suppression mean opinion score (DNSMOS) \cite{Reddy2021}. Subjective speech quality assessment was done using MOS testing. 

\section{Methodology and experimental setup}
\label{sec:method}

\subsection{Datasets}
The UASpeech dataset \cite{Kim2008} is a multimodal 
dysarthric speech dataset 
consisting of words (digits, letters, commands, common and uncommon words). It consists of 15 speakers with dysarthria and 13 speakers with typical  speech. 
The dataset has subjective intelligibility ratings of the dysarthric speakers,
grouping them into \textit{very-low}, \textit{low}, \textit{mid},  and \textit{high} intelligibility levels. Subjects read 3 blocks of words, with each block containing 255 words: 155 words repeated across blocks, and 100 uncommon words that are different across blocks.

The TORGO \cite{rudzicz2012torgo} dataset contains 21 hours of speech collected from 15 speakers: 8 dysarthric speakers with different severity levels (\textit{severe}, \textit{moderate}, \textit{moderate to severe (M/S)} and \textit{mild}, totaling 7.3 hours) and 7 typical speakers (13.7 hours). 
The acoustic data is simultaneously recorded by a head-mounted microphone and an array microphone. TORGO consists of both word and sentence prompts. 

\subsection{Speech enhancement methods}
Noisereduce is a low-complexity signal-processing-based enhancement algorithm which was originally used for denoising animal vocalisations \cite{Sainburg2020}. It has previously been used for removing the stationary noise present in the UASpeech recordings \cite{Harvill2021, Illa2021} and was therefore chosen as the baseline algorithm. 
For our experiments,  we used Noisereduce Version 3 \cite{Sainburg2024} with an FFT length of 512 (corresponding to frames of 32 ms). 

SGMSE, or \textit{score-based generative model for speech enhancement}, is a SE algorithm  based on a stochastic diffusion process which operates in  complex short-time Fourier transform domain \cite{Richter2023}. SGMSE has shown to perform similar to state-of-the-art discriminative SE algorithms in conditions where the training and test sets are matched, while being superior in conditions where the training and test sets were taken from different corpora \cite{Richter2023}. A disadvantage of SGMSE is that it can introduce vocalizing and breathing artifacts in adverse conditions \cite{Richter2023, Lemercier2023}.  We used the publicly available implementation of SGMSE \cite{Welker2023git} with the pre-trained checkpoint on \texttt{VoiceBank/DEMAND}. 

StoRM, or \textit{stochastic regeneration model}, was proposed to lower the computational complexity and to reduce the artifacts associated with SGMSE. StoRM first generates an initial prediction which is subsequently used to guide the diffusion process. We used the publicly available implementation \cite{Lemercier2023git} with the pre-trained checkpoint on \texttt{WSJ0+Chime3} in mode \texttt{storm}. 

We apply the enhancement algorithms to both the dysarthric and the typical speech. This allows the enhanced typical speech to serve as a baseline. 

\subsection{Evaluation Methodology}
Speech intelligibility is objectively measured through ASR performance and speech quality is evaluated through objective and subjective tests.

\subsubsection{Automatic speech recognition performance}

To objectively evaluate the performance of the SE models on the dysarthric speech of both UASpeech and TORGO, we use Whisper-Turbo, the latest version of Whisper--a state-of-the-art model trained on a massive 680k hours of multilingual data \cite{radford2023robust}. Moreover, Whisper-Turbo was FT to investigate the effectiveness of SE on a FT ASR.

For fine-tuning Whisper on UASpeech, we followed \cite{christensen12_interspeech}: blocks 1 and 3 of UASpeech were used for fine-tuning and block 2 for testing. This resulted in 22.91 hours of typical speech and 44.34 hours of dysarthric speech for training and 11.09 hours of typical speech and 21.64 hours of dysarthric speech for testing. Additionally, 10\% of the training data was used for validation during fine-tuning. 
For fine-tuning on TORGO  we used a five-fold cross-validation training strategy, following \cite{yue2020autoencoder}.   The language tag ``en'' was selected during testing. The FT is done with a learning rate of $10^{-5}$ with a step-based evaluation strategy, for $3000$ (UASpeech) and $500$ (TORGO)  training steps. The model uses a linear learning rate scheduler with $100$ warmup steps and a weight decay of $0.01$. Since UASpeech consists only of single words and TORGO consists of a mix of single words and sentences, ASR performance on the typical and dysarthric speech is measured in character error rate (CER) and reported separately for the word and sentence subsets for TORGO. 

\subsubsection{Objective and subjective speech quality}
The objective speech quality is evaluated using DNSMOS \cite{Reddy2021}, which is a no-reference objective speech quality metric used for estimating the MOS of enhanced speech. The DNSMOS score ranges from 1 (bad quality) to 5 (excellent quality). DNSMOS predicts subjective speech quality ratings as would be obtained in a crowd-sourced subjective speech quality study following the ITU-T P.808 standard \cite{Reddy2021, ITUTP808}.
The subjective speech quality assessments were done using MOS testing \cite{Benesty2008}. For each of the 4 intelligibility (UASpeech) / severity levels (TORGO) plus typical speech, and for each of the 3 enhancement methods plus the original data, we randomly selected 6 one-word speech segments (3 male and 3 female speakers; except for the M/S and Moderate speech in TORGO, where there were respectively no female and no male speakers). Thus, there are a total of $5 \times 4 \times 6 =120$ one-word segments selected per database. There were 11 listeners for TORGO (4 female, 6 male, 1 would rather not say; age: 22-32 years) and 12 for UASpeech (3 female, 1 non-binary, 1 male; age 21-34 years), who each rated 80 out of the 120 segments to keep the testing time low (at about 12 minutes). Segments were presented in random order. The participants were instructed to rate the overall speech quality (as a combination of \textit{clarity}, \textit{noisiness}, \textit{distortions} and \textit{naturalness}) on a five-point scale, where the scores $\{1,2,3,4,5\}$ respectively correspond to $\{\text{bad}, \text{poor}, \text{fair}, \text{good}, \text{excellent}\}$ quality. None of the participants had medically diagnosed hearing conditions. All participants were non-native speakers of English, and were recruited through university channels. Participants were not compensated for their participation.

\section{Results}
\label{sec:res}

\subsection{Automatic speech recognition performance}

\subsubsection{Zero-shot testing}
Table \ref{tab:UA+Torgo_ZS} presents Whisper-Turbo's zero-shot testing results for both the original and the enhanced versions of the typical and dysarthric speech of UASpeech and TORGO. As expected, dysarthric speech has a (much) higher CER than typical speech for both datasets.
In general, the CERs for TORGO (Avg) are much lower than those for UASpeech, especially for dysarthric speech. This is partially due to TORGO having better quality speech data than UASpeech. Furthermore,  TORGO contains both sentences and words while UASpeech only has isolated words. Splitting the CERs for TORGO for the word and sentence subsets shows that the CERs for the words subset are indeed much higher than those for the sentences, which benefit from Whisper's strong language modeling. 

For both the UASpeech and TORGO's word subsets, the speech enhanced by Noisereduce gave the best CERs for the dysarthric speech, showing that the SE is beneficial for ASR performance if words are spoken in isolation. This is likely because dysarthric speech often contains irregular articulatory patterns and low-energy phonemes, where SE helps by reducing noise and thereby enhancing weak speech components and key phonetic cues, making the signal more distinguishable for ASR models. The same improvement was not found for sentence-level dysarthric speech, suggesting that a strong language model (LM) outweighs the benefit of SE. Comparing the generative-based SE to the baseline SE we see that the generative-based SE methods do not provide further improvements.
For typical speech, applying SE does not do much or even hurts ASR performance. 
Since typical speech is already highly intelligible, any modification may remove useful spectral details or introduce distortions, leading to a slight degradation in ASR performance. 
This suggests that while SE can aid dysarthric speech recognition in certain conditions, it may not always be beneficial for already intelligible speech.

\begin{table}[t]
  \caption{Results (\%CER) of zero-shot testing Whisper-Turbo on the UASpeech and TORGO datasets. Bold shows the best results across approaches.}
\label{tab:UA+Torgo_ZS}
\centering
\resizebox{\linewidth}{!}{
\begin{tabular}{@{}c|cc|cc|cc|cc}
\toprule
\multicolumn{1}{l|}{}   & \multicolumn{2}{c|}{\textbf{UASpeech}}&   \multicolumn{6}{c}{\textbf{TORGO}}\\
\cmidrule{2-9}
\multicolumn{1}{l|}{}   & \multicolumn{2}{c|}{\textbf{Word}}& \multicolumn{2}{c|}{\textbf{Avg}} &  \multicolumn{2}{c|}{\textbf{Word}} & \multicolumn{2}{c|}{\textbf{Sent}}
\\
\midrule
\multicolumn{1}{l|}{\textbf{}} & \textbf{TYP} &  \textbf{DYS}  &\textbf{TYP}    & \textbf{DYS}        & \textbf{TYP}       & \textbf{DYS}        & \textbf{TYP}         & \textbf{DYS}        \\
\midrule
\textbf{Original}      & \textbf{9.3} & 165.2 & 5.3 & \textbf{39.1} & 18.2 & 85.6  & \textbf{1.3} & \textbf{22.7} \\ 
\textbf{Noisered.} & 9.9 & \textbf{104.5} & \textbf{5.1} & 48.3 &\textbf{16.8} & \textbf{77.2} & 1.4 & 38.2 \\ 
\textbf{StoRM} & 10.5 & 115.2 & 7.2 & 53.0& 24.5 & 89.7 & 1.6 & 40.1  \\
\textbf{SGMSE}  & 11.2 & 133.6 &5.5 & 51.9& 17.9 & 83.1 & 1.6 & 40.9 \\
\bottomrule
\end{tabular}
}
\end{table}

\subsubsection{The effect of SE on fine-tuning Whisper}
Table \ref{tab:UA+Torgo_FT} shows the results of the ASR performance after fine-tuning Whisper on the original and the enhanced versions of the typical and dysarthric speech of UASpeech and TORGO.

\begin{table}[ht]
  \caption{Results (\%CER) of fine-tuning Whisper-Turbo on UASpeech and TORGO, for the different intelligibility/severity level separately. Bold shows best results across approaches.}
\label{tab:UA+Torgo_FT}
\centering
\resizebox{\linewidth}{!}{
\begin{tabular}{l|c|cc|cccc}
\toprule
\multicolumn{2}{l|}{}   &   \multicolumn{2}{c}{} & \multicolumn{4}{c}{\textbf{DYS: Intelligibility}}              \\ \midrule

 \multicolumn{2}{c|}{\textbf{Fine-tuning}} & \textbf{TYP} &  \textbf{DYS}        & \textbf{V. low} & \textbf{Low} & \textbf{Mid} & \textbf{High}   \\
\midrule
\multirow{4}{*}{\rotatebox{90}{\textbf{UASpeech}}} & \textbf{Original}        & 25.9 & \textbf{43.0} & \textbf{64.4}	& \textbf{44.8} &	\textbf{40.1}	& 30.2  \\ 
& \textbf{Noisered.} &   24.5	& 43.6	& 65.3	& 45.7	& 41.1	 & 30.1 \\ 
& \textbf{StoRM}  & \textbf{22.3} &	45.5 & 69.3 & 48.3 &	43.3	& \textbf{28.7 }  \\
& \textbf{SGMSE}  & 26.0	& 47.3 & 70.7 & 49.7	& 45.8 &	32.0   \\
\midrule
\multicolumn{2}{l|}{}   &   \multicolumn{2}{c}{\textbf{}} & \multicolumn{4}{c}{\textbf{DYS: Severity}} \\
\midrule
  \multicolumn{2}{c|}{\textbf{Fine-tuning}} & \textbf{TYP}         & \textbf{DYS}         & \textbf{Severe} & \textbf{M/S} & \textbf{Mod.} & \textbf{Mild} \\
\midrule
\multirow{4}{*}{\rotatebox{90}{\textbf{TORGO}}} & \textbf{Original}        &  5.2   &  \textbf{ 19.3}    & \textbf{ 30.0 }   &   18.1   & 15.1
& 5.9 \\ 
& \textbf{Noisered.} &  5.4   &   20.1   & 31.3 &   21.1        
&  15.0 & 6.1 \\ 
& \textbf{StoRM} & 6.4 &   23.4    &  38.4    &    \textbf{17.7}  
& 17.2 &  6.4 \\
& \textbf{SGMSE}  & \textbf{5.1}    &  19.9   &  32.4   & 17.9
& \textbf{14.2}& \textbf{5.3}\\
\bottomrule
\end{tabular}
}
\end{table}

After FT on the original typical and dysarthric speech, the ASR performance on the dysarthric speech is greatly improved for both datasets (compare the row ``original'' with the results in Table \ref{tab:UA+Torgo_ZS}), while for typical speech the performance degraded substantially for UASpeech and stayed similar for TORGO.

After FT on the enhanced typical and dysarthric speech, in general there is no further performance improvement of the dysarthric ASR. One possible reason is that once the model is fine-tuned on dysarthric speech, additional enhancement of the dysarthric speech is no longer beneficial. Thus, (generative) SE has no additional benefit to fine-tuning. However, SE does improve FT performance for dysarthric speech that is less severe (TORGO) or highly intelligible (UASpeech). Comparing across different severity levels, for high severity level speech (with very low intelligibility) the original speech has the lowest CER, while for dysarthric speakers with very low to mid intelligibility, the diffusion-based enhanced speech performs the best (28.7\% for StoRM-enhanced UASpeech and 5.3\% for SGMSE-enhanced TORGO). 

We hypothesize that, in zero-shot testing, SE helps for word-level ASR performance by reducing noise, which is beneficial when no domain-specific speech is available to train the ASR system. 
When fine-tuning is possible however, the ASR system adapts to the noisy and dysarthric speech, leading to a recognition results that well outperforms the advantage of (generative) SE. Moreover, generative SE may introduce phonetic confusions or other `speech-like' artifacts, causing information loss \cite{Richter2023}. We speculate that this effect is more pronounced for lower intelligibility speakers whose speech has more distortions, explaining why the typical and high intelligibility dysarthric speakers can still benefit from StoRM or SGMSE.

\subsection{Speech quality} 
\subsubsection{Objective speech quality}
\begin{figure}[ht]
\centering
\includegraphics[width=1\columnwidth, trim={1.4cm 1.2cm 1.2cm 1.2cm}, clip]{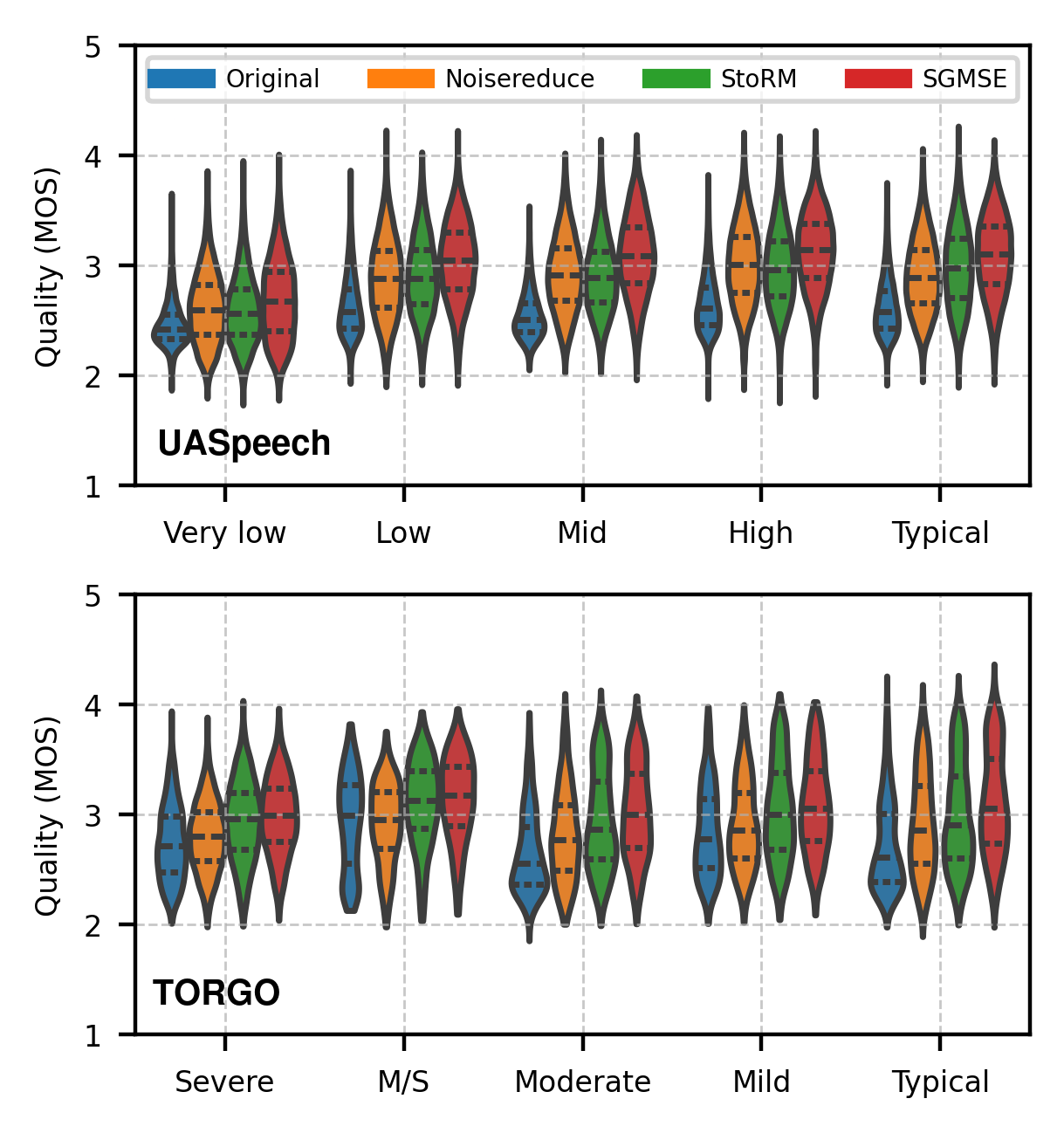}
    \caption{{Violinplots showing the DNSMOS results for objective speech quality for UASpeech and TORGO, expressed as MOS, ranging from 1 to 5 (bad to excellent quality). Dashed lines indicate the quartiles. Ordered by intelligibility of dysarthric speakers (UASpeech) and  severity of dysarthria (TORGO).}}  
    \label{fig_obj_dnsmos}
\end{figure}

The objective speech quality results, measured using DNSMOS, are shown in Fig. \ref{fig_obj_dnsmos} using violinplots. The violinplots visualize the distribution of the speech quality datapoints. The width of the violin indicates the concentration of datapoints around that quality level, where a larger width indicates a higher concentration. The quartiles are indicated by the dashed lines and the scores range from 1 to 5, where higher is better. All three SE algorithms improve the objective speech quality compared to the original data (in blue), as can be seen by the violins moving upwards. 
Of the SE algorithms, SGMSE attains the highest speech quality scores, as can be seen by the datapoints being more concentrated around higher MOS score. SGMSE is followed closely by StoRM and Noisereduce. Additionally, for UASpeech (Fig. \ref{fig_obj_dnsmos}, top panel) there is an increase in speech quality with increasing intelligibility of the dysarthric speaker. This is likely due to the increasing deviation of dysarthric speech to typical speech as the intelligibility decreases. This effect is not as visible for TORGO  (Fig. \ref{fig_obj_dnsmos}, bottom panel). 
For TORGO, it can also be seen that the quality scores are more dispersed compared to the results for UASpeech, and sometimes with two different wide spreads. This can be attributed to the two types of microphone used in TORGO: an array and a head-mounted microphone \cite{rudzicz2012torgo}. Inspecting the results separately per microphone type (not plotted due to space constraints), it is found that the speech quality for the microphone array is lower compared to that of the head-mounted microphone, leading to a wider spread in speech quality values. 

In conclusion, speech enhancement of dysarthric speech improves the objective quality of the dysarthric speech, which seems to be somewhat larger for speech that is closer to typical speech. 

\subsubsection{Subjective speech quality}
The subjective speech quality results are presented per dysarthric intelligibility/severity level in Table \ref{tab:mos_uaspeech_torgo} for UASpeech (top) and TORGO (bottom). Table \ref{tab:mos_uaspeech_torgo} shows that speech quality for high intelligibility/mild severity speakers is perceived to be close to that of typical speech (in line with earlier results on UASpeech \cite{Illa2021}), and drops for increasingly reduced intelligibility/increased severity of the dysarthria (where higher is better). Overall, applying SE increases the perceived speech quality for all dysarthria levels and both databases. Except for one case (UASpeech, low intelligibility), the generative-based SE SGMSE is consistently rated as having the best speech quality for all dysarthric speech severity levels and typical speech, which is in line with the objective results. 
The subjective data shows no clear preference for StoRM over Noisereduce or vice versa.

\begin{table}[h]
  \caption{Subjective MOS results for UASpeech and TORGO. Mean and standard deviation are indicated ($\mu\pm\sigma$). Results range from 1 to 5 (bad to excellent quality). Bold shows best results across approaches.}
  \label{tab:mos_uaspeech_torgo}
\centering
\resizebox{\linewidth}{!}{
\begin{tabular}{@{}c|ccccc@{}}
\toprule
    \textbf{UASpeech} &\textbf{Very Low} & \textbf{Low} & \textbf{Mid} & \textbf{High} & \textbf{Typical} \\
\cmidrule{1-6}
Original     & $1.7 \pm 0.8$ & $2.5 \pm 1.1$ & $2.4 \pm 1.1$  & $3.3 \pm 1.1$  & $3.2 \pm 1.0$  \\
Noisered.    & $2.1 \pm 0.9$ & $\mathbf{3.2} \pm 1.2$ & $2.9 \pm 1.1$  & $4.0 \pm 0.9$  & $4.0 \pm 0.8$  \\
StoRM        & $2.1 \pm 1.0$ & $2.9 \pm 1.1$ & $2.7 \pm 1.2$  & $3.8 \pm 0.8$  & $\mathbf{4.4} \pm 0.7$  \\
SGMSE        & $\mathbf{2.3} \pm 1.3$ & $3.1 \pm 1.3$ & $\mathbf{3.0} \pm 1.2$  & $\mathbf{4.1} \pm 1.0$  & $\mathbf{4.4} \pm 0.9$  \\
\bottomrule
\toprule
\textbf{TORGO} &\textbf{Severe} & \textbf{M/S} & \textbf{Moderate} & \textbf{Mild} & \textbf{Typical} \\
\cmidrule{1-6}   
Original     & $2.4 \pm 1.2$ & $2.1 \pm 1.3$ & $2.4 \pm 1.1$  & $3.7 \pm 1.0$  & $3.1 \pm 1.3$  \\
Noisered.    & $2.7 \pm 1.2$ & $2.5 \pm 1.5$ & $2.8 \pm 1.3$  & $4.5 \pm 0.7$  & $3.9 \pm 1.0$  \\
StoRM        & $2.3 \pm 1.0$ & $2.7 \pm 1.3$ & $3.1 \pm 1.0$  & $4.4 \pm 0.7$  & $3.7 \pm 1.1$  \\
SGMSE        & $\mathbf{3.1} \pm 1.3$ & $\mathbf{3.2} \pm 1.4$ & $\mathbf{4.0} \pm 1.0$  & $\mathbf{4.8} \pm 0.4$  & $\mathbf{4.5} \pm 0.8$  \\
\bottomrule
\end{tabular}
}
\end{table}

\section{Discussion and Conclusion}
In this paper, we investigated the effectiveness of diffusion-based and signal-processing-based speech enhancement (SE) methods on two commonly used dysarthric speech datasets -- UASpeech and TORGO. We evaluated the resulting speech intelligibility with Whisper-Turbo and the resulting speech quality both objectively (DNSMOS) and subjectively (through MOS testing).  

Zero-shot testing Whisper-Turbo on the original and enhanced dysarthric and typical speech showed that while, SE has a positive effect on dysarthric speech, it might degrade the performance of the already well-intelligible typical speech. Fine-tuning Whisper-Turbo improves speech recognition performance for the dysarthric speakers, but not for the typical speakers. This is the case for both the enhanced and the original speech. However, comparing the performance after fine-tuning between the original and enhanced speech, it is found that the ASR performance on the enhanced dysarthric speech is lower than that on the original speech. This is in particular the case for speakers with severe dysarthria and with low intelligibility. For high intelligibility dysarthric speakers and for speakers with moderate and mild dysarthria, SE can still have benefits.

Objective and subjective testing of speech quality showed that SE improved the speech quality of the dysarthric and typical speech, likely by removing the stationary noise present in the data. Generative SE achieved the largest quality improvement, with SGSME outperforming the other speech enhancement methods. 

Comparing the ASR results with the objective and subjective speech quality results shows that while SE consistently improves both objective and subjective speech quality, this does not always translate to better ASR performance -- only for dysarthric speech spoken in isolation, and only for signal processing-based SE. Potentially, while SE improves speech quality by reducing stationary noise, its impact on intelligibility is more complex. In cases where SE distorts critical speech cues, particularly for severely dysarthric speakers, the enhancement may degrade ASR performance despite perceived quality improvements. This is especially the case after fine-tuning and for generative SE. 

 Our research was inspired by \cite{Reszka2024}, who found that using generative diffusion-based SE on dysarthric speech improved dysarthric speech detection. Our experiments on ASR show that the benefit of generative diffusion-based SE does not extend to ASR. \cite{Reszka2024} hypothesised that generative diffusion-based SE moved the distribution of the dysarthric speech closer to that of typical speech by removing acoustic dysarthric speech markers. Since our results showed a clear improvement in terms of objective and subjective speech quality, potentially this shift in distribution of the speech samples results in improved speech quality but removes important speech markers for ASR. It is likely that the majority of the increase in quality is due to removing the stationary noise of the speech in UASpeech and TORGO. 

In future work, SE will be investigated as a data augmentation approach for dysarthric speech recognition. Additionally, we plan to further investigate the effect of SE on dysarthric speech quality and ASR by computing the Kullback-Leibler divergence between the (diffusion-based) enhanced dysarthric speech and typical speech. Since there are considerable challenges in estimating these distributions correctly, it was considered to be beyond the scope of this work. We will also use language-model free models such as Wav2Vec 2.0.  

In conclusion, speech enhancement of dysarthric speech does not provide additional benefits when using a strong ASR model, when a strong language model can be used (sentences vs. words results for TORGO) or when the ASR can be fine-tuned. However, when one wants to recognize isolated words using an off-the-shelf ASR system, i.e., without retraining or fine-tuning, signal processing-based SE may yield improved recognition performance. Moreover, different (possibly lower-complexity) ASRs might still benefit from SE.

\clearpage

\bibliographystyle{IEEEtran}
\bibliography{mybib}

\end{document}